\documentclass[a4paper,11pt]{article}
\usepackage[latin1]{inputenc}
\usepackage[T1]{fontenc}
\usepackage[english]{babel}
\usepackage{latexsym}
\usepackage{graphicx}
\usepackage{wrapfig}
\addto\captionsenglish{}
\begin{document}
\title{An easy journey from Galilean to General Relativity}
\author{Stéphane Fay\footnote{steph.fay@gmail.com}\\
Palais de la Découverte\\
Astronomy Department\\
Avenue Franklin Roosevelt\\
75008 Paris\\
France
}
\maketitle
\begin{abstract}
We explain in a very concise way the basic principles that lead from Galilean to General Relativity to make them understandable to students or general audience, even with little knowledge in physics and mathematics.
\end{abstract}
\section{Introduction}
This paper explains the basic principles that lead from Galilean to General Relativity in a simple and concise way to make them understandable in a few pages, even for students or general audience with little knowledge in physics and mathematics. To do so, we do not use any calculation (other than the Pythagorean theorem and some plane geometry ideas) and we do not try to be extensive (for example, we do not explain the famous formula $E = mc^2$ or all the evidences in favor of General Relativity). We rather follow a sequence of concepts that starts with the Galilean Relativity principle and its velocity addition law. Noting that this law does not work for light, we are led to the generalization of the Galilean Relativity into Special Relativity and its space-time concept. An example of space-time, analogous to the plane geometry of Euclid, is presented. It allows to address the idea of distance between events in space-time. We then define the gravitational and inertial masses which lead to the equivalence principle. It enable to understand why $100$ grams of feathers fall the same way as $1$ kilogram of lead. Through this principle, it is possible to establish a link between Special and General Relativity. We also explain why this last theory is a gravitation theory by showing how it implies that matter curves space-time. We then present the advance of the Mercury perihelion as one of the proof of General Relativity. We conclude on the status of General Relativity today. Note that this paper does not contain an extensive bibliography on Relativity, that should then be larger than the paper itself, but only some few historical references.
\section{Galilean Relativity}
In the early seventeenth century, the Italian astronomer Galileo Galilei studies the relativity of motion. He imagines an experiment\cite{{Gal32}} in which an observer drops a ball from the top of the mast of a boat sailing along a straight line, at constant velocity. 

\begin{wrapfigure}[10]{r}{0.5\textwidth}
\includegraphics[width=0.50\textwidth]{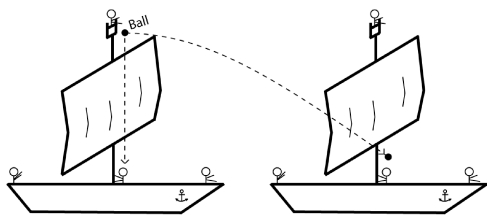}
\caption{\scriptsize{\label{fig0}Observers on the boat see the ball falling vertically but not observers on the docks.}}
\end{wrapfigure}
\noindent He remarks that this observer always sees the ball to fall vertically at the foot of the mast. However, for a stationary observer on the docks looking at the boat, the ball does not describe a straight trajectory but a curved one combining two motions, the fall of the ball along the mast and the race of the boat as shown on figure \ref{fig0}. The motion of the ball is thus relative to a given observer. However, whatever the observer, the velocity $v$ of the ball is always the sum of two relative velocities with respect to this observer, the boat velocity $v_1$ and the ball velocity $v_2$. It is the velocity addition law. If the observers on the docks and on the boat do not perceive the same trajectory for the ball, it is because for the observer on the docks, the boat velocity $v_1\not =0$ whereas for the observer on the boat, $v_1 = 0$.\\
The latter equality means that an observer on a boat going in a straight line at a constant velocity does not feel the motion of the boat. In particular, if he is shut in a ship cabin with no way to see the outside, no experiment with a ball or any physical body allows him to determine if the boat is docked or sailing on a quiet sea. The result of an experiment in mechanics, the branch of science concerned with the behavior of physical bodies subject to forces or displacements, is thus independent from the motion of the observer when he moves in a straight line at constant velocity: this is the Galilean Relativity principle.
\section{Special Relativity}
\begin{wrapfigure}[10]{r}{0.5\textwidth}
\includegraphics[width=0.60\textwidth]{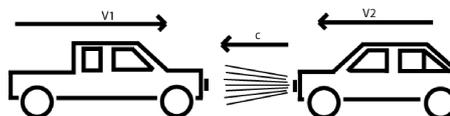}
\caption{\scriptsize{\label{fig2}Both cars pass each other at the relative velocity $v_1 + v_2$. Light of the second car propagates from it at speed $c$. However, it does not reach the first car at the velocity $v_1 + v_2 + c$ but always at speed $c$, whatever $v_1$ and $v_2$.}}
\end{wrapfigure}

In $1676$, the Danish astronomer Ole Christensen Rømer discovers that light has a finite speed\cite{Rom76}, denoted $c$. Today we know that $c = 299,792,456$ meters/second in vacuum. This constant value for the speed of light is a problem for the velocity addition law of the Galilean Relativity principle. Let us explain why. Imagine that you are in a car moving at velocity $v_1$. A second car comes toward you at velocity $v_2$. The velocity addition law implies that both cars pass each other at the relative velocity $v_1 + v_2$ as shown on \ref{fig2}.
But if the second car turns on its headlights, the light does not reach you at the velocity $v_1 + v_2 + c$ but always at speed $c$: the velocity addition law therefore does not work with light! The reason is that this law is, indeed, an approximation, valid only for objects moving at small velocities compared to that of light. In other words, the Galilean Relativity and its velocity addition law do not apply for velocity close to $c$, and thus do not apply to light.\\
In $1905$, the German physicist Albert Einstein solves this problem by extending the Galilean Relativity principle\cite{Ein05}. He postulates that all the laws of physics, not only the laws of mechanics, are the same for any observer at rest or moving in a straight line at constant velocity. This is the Special Relativity principle that applies to matter but also to light. In particular, the finite speed value $c$ of light is erected as a physical law: it has always the same value for all the observers at rest or moving in a straight line at constant velocity with respect to the light source.\\
But then, how can we explain the constancy of $c$ despite the relative motion of an observer with respect to a light source? The speed of light is, like any velocity, the ratio of a distance $x$ to a time $t$, $c = x / t$. So that $c$ remains constant when an observer is moving, distance and time have thus to vary together. This means that space and time are related and not independent absolute entities: this is the birth of the space-time concept.
\section{Space-time}
Space-time is not mysterious at all. When you make an appointment, you choose a precise point in the three-dimensional space and a specific time. So every day you use four coordinates to locate yourself in space-time.\\
Usually, one considers space and time as two separate entities: then, moving in space does not change the way you "move" in time. This is a good approximation in everyday life, for small velocities. However it does not work when an observer moves at nearly the speed of light. Then Special Relativity applies, and the spatial distance and time interval are not absolute but related to each other. This can be seen through two effects that have been experimentally verified: length contraction and time dilation.\\
Length contraction implies that, for a stationary observer, the length of an object moving at high velocity should seem to contract in the direction of its motion\footnote{Note that this phenomenon is usually not directly observable due to optical effects\cite{Ter59}.}. Time dilatation implies that if you travel in a very fast spaceship, the travel time as measured on Earth will be longer than the one you measure in the spaceship. Hence, moving in space changes the way you "move" in time with respect to a stationary observer: time and space are thus related to each other.
\section{A simple example of space-time}\label{s5}
\begin{wrapfigure}[15]{r}{0.4\textwidth}
\includegraphics[width=0.4\textwidth]{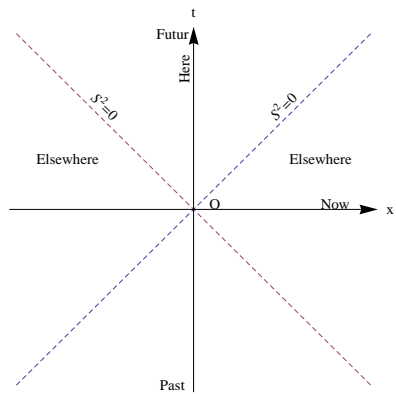}
\caption{\scriptsize{\label{fig1}Minkowski space-time.}}
\end{wrapfigure}

To make more tangible the concept of space-time, let us plot one of them. We can represent space-time by making an analogy with plane geometry. In plane geometry, we have two spatial dimensions. We thus use two space coordinates $x$ and $y$ to identify a point in the plane. The distance $D$ that separates two points of coordinates $(x_1, y_1)$ and $(x_2, y_2)$ satisfies: $D^2 = (x_2-x_1)^2+ (y_2-y_1)^2$. This is the Pythagorean theorem.\\
In Special Relativity, time and space are combined. To determine a point in a two dimensional space-time, we use one space $x$ and one time $t$ coordinates. The analogue of plane geometry for space-time geometry is the Minkowski geometry\cite{Min08} that is represented on figure \ref{fig1}.

\noindent Each point $(x_i,t_i)$ of this space-time is called an "event" because it takes place in a particular place $x=x_i$ and time $t=t_i$. $O$ is an observer here $(x = 0)$ and now $(t = 0)$. The two dashed lines leaving $O$ (to the top) represent light rays going to his future $(t> 0)$. The two dashed lines arriving in $O$ (from the bottom) represent light rays reaching him from his past $(t< 0)$. The fact that $O$ cannot move faster than light implies that he can only reach events in the future between the two dashed line with $t> 0$. Similarly, all the events located in the past of $O$ and that can act with him are between the two dashed lines with $t <0$. The events outside of these lines are "elsewhere". They cannot reach or be reached by $O$ without exceeding the speed of light.\\
As in plane geometry for which the Pythagorean theorem defines the distance between two points in space, the distance $S$ between two events of the Minkowski space-time separated by a distance $x$ and a time $t$ is defined as $S^2 = c^2t^2-x^2$. While the measurements of $t$ and $x$ are relative to the motion of the observer, the measurement of $S^2$ is independent from it. It is thus an absolute value. Since the speed of light is $c = x / t$, we reach the important conclusion that the distance between two events separated by a light ray is always $S^2 = 0$. This means that light always takes the shortest path between two events in space-time. This shortest path is called a "geodesic". In plane geometry, it is a straight line.
\section{Equivalence principle}
The principles of Galilean Relativity and Special Relativity only consider motions in a straight line at constant speed. In $1915$, Einstein succeeds to also consider the accelerated observers thanks to the so-called equivalence principle. This gives birth to General Relativity.\\
To understand the equivalence principle, we need to define the concept of mass in two ways. First, the "gravitational mass" $m_p$ determines the force experienced by a body in a gravitational field. Second, the "inertial mass" $m_i$ determines the resistance of a body to a change of motion under the effect of a force. Thus, when a mass falls on Earth, gravity acts on it with a force $F = m_p g$ where $g = 9.81 m/s^2$ is the intensity of the gravity field for our planet. But the same force $F$ is required to overcome the inertia of the mass and thus it also writes $F = m_i a$, with "a" the acceleration of the mass to the ground. A physical principle called the weak equivalence principle states that $m_p = m_i$. It is experimentally verified at high precision level. Comparing the two above expression for $F$, equality between gravitational and inertial masses thus implies that $a = g$. 
\begin{wrapfigure}[11]{r}{0.5\textwidth}
\includegraphics[width=0.50\textwidth]{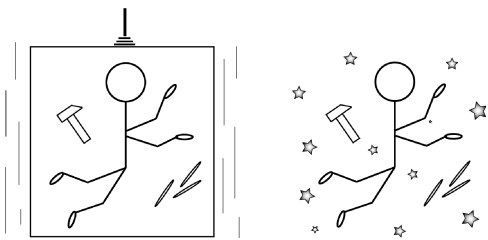}
\caption{\scriptsize{\label{fig3}You cannot know if you are in a falling elevator (left) or weightless in space (right): in both cases, you float !}}
\end{wrapfigure}
\noindent Hence, any object falling to the ground because of gravity, falls with the same acceleration, whatever its mass and composition: $100$ grams of feather thus fall the same way as $1$ kilogram of lead! Of course, this is true in vacuum. In presence of air, the friction implies that the feather falls slower than the lead. Now imagine yourself in a falling elevator with a feather and a hammer. All these items falling in the same way, including yourself, nothing seems to move in the elevator: you cannot know if you are falling on Earth or weightless in space as illustrated on figure \ref{fig3}. This "thought experiment" establishes the equivalence between free fall and weightlessness, or between acceleration and gravity: this is the Einstein equivalence principle. It allows to make a link between Special Relativity and General Relativity. How ?\\
In Special Relativity, the laws of physics are the same for any observer moving in a straight line at constant speed in vacuum. General Relativity tells us that the laws of physics are the same for any observer, even if he is accelerating in the vacuum. The link between these two theories is established through the Einstein equivalence principle. This last one implies that gravity can be canceled locally by an acceleration, as in a falling elevator. Hence, in the elevator accelerated by Earth gravity where General Relativity applies, everything happens as if you were floating in space, without any acceleration, and where Special Relativity applies. The laws of physics are thus the same for any observer, his motion being accelerated as in General Relativity or not accelerated as in Special Relativity. A constant velocity motion being a special case of an accelerated motion, Special Relativity is thus a special case of General Relativity.
\section{General Relativity}
The link between gravity field and acceleration implies that General Relativity is also a gravitation theory. This becomes obvious when one understands how mass curves space-time\cite{Ein15}.

\begin{wrapfigure}[12]{r}{0.5\textwidth}
\includegraphics[width=0.50\textwidth]{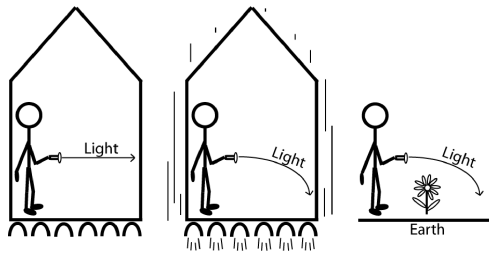}
\caption{\scriptsize{\label{fig4}Acceleration and gravitation are equivalent. Hence a light ray is curved by rocket acceleration as well as Earth gravity.}}
\end{wrapfigure}
\begin{wrapfigure}{r}{0.5\textwidth}
\includegraphics[width=0.40\textwidth]{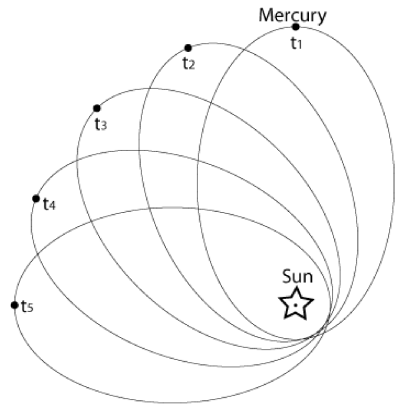}
\caption{\scriptsize{\label{fig5}Mercury orbit changes its orientation with time $t_i$. Hence, the point of closest approach of Mercury to the Sun, called the perihelion, slowly moves around the Sun.}}
\end{wrapfigure}

\noindent Let us imagine a stationary rocket in the vacuum of space. Inside the rocket, an astronaut turns on a lamp: the light ray then propagates along a straight line in the rocket like on the first picture of figure \ref{fig4}. Now let us assume that the rocket is accelerating. Once again, an astronaut turns on a lamp. Initially, the light ray starts propagating at a height $h$ with respect to the rocket ground. But a moment later, as the rocket is accelerating, the light ray is nearer from the ground: it thus follows a curved path with respect to the rocket ground like on the second picture of figure \ref{fig4}. Now the Einstein equivalence principle implies that acceleration and gravity are the same. Therefore, if our astronaut turns on his lamp on Earth, the light ray will also be bent like in the accelerated rocket, but this time because of Earth gravity like on the third picture of figure \ref{fig4}. But we have said at the end of section \ref{s5} that a light ray always follows the shortest path through space-time (in vacuum). If this shortest path is curved, it is that space-time itself is curved. Hence, gravity generated by a massive body curves space-time and General Relativity is thus a gravitation theory.\\
The concept of gravity according to Einstein General Relativity is then very different from that of the English physicist and mathematician Isaac Newton. For this latter, gravity is a force: if the Earth revolves around the Sun, it is because our star exerts a force on our planet that allows it to rotate around it without moving away. In General Relativity, if the Earth revolves around the Sun, it is not because it is attracted by a gravitational force but because it goes straight in a space-time curved by the mass of the star. Similarly, on a velodrome track, cyclists follow the curvature of the track without turning their handlebars and, despite that, run a closed path like the Earth around the Sun.\\
The first evidence in favor of General Relativity was given by the trajectory of the planet Mercury. A planet describes an ellipse with the Sun at one of its focus. The perihelion is the closest point from the Sun on this path. It is found that this point moves slightly with time as shown on figure \ref{fig5}. 

\noindent This phenomenon is called precession. It is due to the gravitational influence of other planets. In the case of the planet Mercury, there is a small difference of $42.98$ arc-seconds\footnote{$1$ arc-seconds $=1/3600$ of a degree} per century between the precession predicted by Newton's theory and the observed one. This was noticed in $1859$ by the French astronomer and mathematician Urbain Le Verrier\cite{Ver59}. Some had then imagined the presence of a new planet named Vulcan, closer to the Sun than Mercury, which would have perturbed the trajectory of the latter. But no one has ever found Vulcan and finally Newton's theory of gravity was questioned. Hence, the 18 November 1915, Albert Einstein\cite{Ein15A} reported to the Prussian Academy that, contrary to Newton's theory of gravity, General Relativity with its concept of curved space-time predicted precisely the observed precession of Mercury perihelion. That was the first triumph of Einstein General Relativity. Let us remark that it does not mean that Newton's theory of gravity is "false". Indeed a scientific theory is never "true" or "false". It is always an approximation of Nature's law. In the case of Newton's theory, it is a valid approximation to describe gravitation around objects with small masses in a similar way that Galilean principle is valid for objects with small velocities.
\section{Conclusion}
What is the status of General Relativity today? It is interesting to note that it suffers, among others, a similar problem as that of Newton theory with Mercury. Observing the trajectories of stars in some galaxies, the American astronomer Vera Rubin showed in the seventies that they are not exactly those predicted by General Relativity\cite{Rub70}! This problem can be solved in two ways: ever we consider the presence of an invisible matter (the so-called dark matter) which would disturb the trajectories of stars in their galaxies or we modify the Einstein gravitational theory. It is still unclear what is the right choice. But in any case, we already know that General Relativity is not sufficient to study the most extreme objects of the Universe such as black holes. General Relativity will thus have to evolve, one way being to merge it with Quantum Mechanics describing the world of particles to form a Quantum Gravity theory.
\section*{Acknowledgements}
I thank Romain Attal for long and enlightening discussions. This paper is the outcome of an exhibition on Einstein General Relativity that took place in "Palais de la Découverte" (Paris) in September $2015$.
\bibliographystyle{unsrt}

\end{document}